\documentclass[aps,prl,reprint,twocolumn,superscriptaddress,floatfix,nofootinbib,longbibliography,preprintnumbers]{revtex4-1}
\usepackage{amsthm}
\usepackage{amsmath,amssymb, color, comment, physics}
\usepackage[makeroom]{cancel}
\usepackage[caption=false]{subfig}
\usepackage{mathrsfs}
\usepackage{graphicx}
\usepackage{subfig}
\usepackage[countmax]{subfloat}
\usepackage[english]{babel}
\usepackage{dsfont}
\usepackage{bm}
\usepackage[bookmarks=true,colorlinks,linkcolor=OrangeRed,urlcolor=NavyBlue,citecolor=RoyalBlue]{hyperref}
\usepackage[dvipsnames]{xcolor}
\usepackage{braket}
\usepackage{mathtools}
\usepackage{soul} 
\usepackage{comment}
\usepackage{ulem}

\definecolor{mygreen}{rgb}{0.25,0.5,0.25}

\begin{document}
%TC:ignore
\preprint{RBI-ThPhys-2025-39}

\title{Experimental preparation of $W$ states through frustration on a programmable quantum simulator}
%\title{Engineering $W$ states from topological frustration in a programmable quantum simulator}
%\title{Frustration induced $W$ states in a programmable Rydberg quantum simulator}
%\title{Engineering $W$ states through frustration on a programmable Rydberg quantum simulator}

\author{Alberto Giuseppe Catalano$^+$}
\email{albertogiuseppe.catalano@unipd.it}
\affiliation{Dipartimento di Fisica e Astronomia "G. Galilei" \& Padua Quantum Technologies Research Center, Universit{\`a} degli Studi di Padova, Italy I-35131, Padova, Italy}
\affiliation{INFN, Sezione di Padova, via Marzolo 8, I-35131, Padova, Italy}
\affiliation{Institut Ru\dj er Bo\v{s}kovi\'c, Bijeni\v{c}ka cesta 54, 10000 Zagreb, Croatia}

\author{Ceren B.~Da\u{g}$^+$}
\email{ceren_dag@g.harvard.edu}
\affiliation{Department of Physics, Harvard University, 17 Oxford Street Cambridge, MA 02138, USA}
\affiliation{ITAMP, Harvard-Smithsonian Center for Astrophysics, Cambridge, Massachusetts, 02138, USA}
\affiliation{Department of Physics, Indiana University, Bloomington, Indiana 47405, USA}

\author{Gianpaolo Torre}
%\email{gianpaolo.torre@irb.hr}
\affiliation{Institut Ru\dj er Bo\v{s}kovi\'c, Bijeni\v{c}ka cesta 54, 10000 Zagreb, Croatia}
\affiliation{Institut Jo\v{z}ef Stefan, Jamova cesta 39, 1000, Ljubljana, Slovenia}

\author{Salvatore Marco Giampaolo}
%\email{sgiampa@irb.hr}
\affiliation{Institut Ru\dj er Bo\v{s}kovi\'c, Bijeni\v{c}ka cesta 54, 10000 Zagreb, Croatia}

\author{Fabio Franchini}
\email{fabio@irb.hr}
\affiliation{Institut Ru\dj er Bo\v{s}kovi\'c, Bijeni\v{c}ka cesta 54, 10000 Zagreb, Croatia}

\begin{abstract}
$W$ states are a central class of multipartite entangled states with applications in quantum information processing, yet their scalable and deterministic preparation remains challenging. 
Here we propose a protocol based on {\it topological ring frustration}, where an antiferromagnetic ring with an odd number of sites hosts a delocalized excitation corresponding to a $W$ state. 
We implement this protocol on a Rydberg atom array --a programmable quantum simulator-- generating $W$ states of up to 11 atoms. 
Our results demonstrate a fidelity of $\mathcal{F} \approx 0.77$, and numerical simulations indicate scalability to larger system sizes accessible with near-term hardware improvements. 
To enable certification of these many-body entangled states, we introduce a novel and efficient Bayesian tomography method that, leveraging on classical simulations, enables their certification with a cost that avoids the exponential scaling of full tomography. 
These results establish topological frustration as a practical mechanism for engineering multipartite entanglement and provide a scalable route toward the certification of correlated quantum many-body states in quantum simulators.
\end{abstract}

\maketitle

\def\thefootnote{+}\footnotetext{These authors contributed equally to this work.}
%TC:endignore
\section{Introduction}
Understanding which quantum resources enable quantum advantage remains a central challenge in quantum science~\cite{Chitambar2019}. 
Several candidates have been proposed, including entanglement~\cite{Plenio2007, Horodecki2009}, discord~\cite{Ollivier2001, Modi2012}, coherence~\cite{Baumgratz2014, Streltsov2017}, and non-stabilizerness~\cite{Howard2017, Leone2022}, yet no single resource provides a unified explanation across all computational and information processing tasks. 
Hence, it is arguably more important to focus on preparing specific quantum entangled states that are known to play a key role in known protocols and algorithms.

A prominent example is the $W$ state, in which a single excitation is symmetrically delocalized across all subsystems~\cite{Dur2000}, i.e.,
\begin{equation}
    \label{eq:Wstatedef}
    \ket{\rm W}  \equiv \frac{1}{\sqrt{L}} \Big( \ket{10\ldots0} + \ket{01\ldots0} + \ldots + \ket{00\ldots 1} \Big) \; .
\end{equation}
Together with the Greenberger–Horne–Zeilinger (GHZ) state~\cite{Greenberger1989}, expressed as 
\begin{equation}
    \label{eq:GHZdef}
    \ket{\rm GHZ} \equiv \frac{1}{\sqrt{2}} \Big( \ket{00\ldots0} + \ket{11\ldots1} \Big) \; ,
\end{equation}
they represent the two inequivalent classes of genuine multipartite entanglement in three-qubit systems. 
On one hand, GHZ-like states possess purely multipartite entanglement that is highly fragile, e.g., vanishing upon the loss or measurement of a single qubit. In contrast, $W$ states combine multipartite and bipartite correlations, retaining nontrivial entanglement even when one qubit is lost. 
This resilience, together with their richer correlation structure, makes $W$ states particularly relevant for quantum information processing tasks, with a broad range of applications to quantum communication protocols, including teleportation of unknown states~\cite{Joo2003}, quantum secret sharing and key distribution~\cite{Wang2007, chen2008}, random number generation~\cite{grafe2014}, quantum thermodynamics~\cite{daug2016}, and photonic and atomic encoding schemes~\cite{Ebert2015, Vijayan2020}.
Furthermore, they provide the fundamental initial resource in Grover-like quantum searches~\cite{Grover1997} restricted to specific subspaces~\cite{Biham2002, Shapira2005}. This importance has motivated extensive efforts toward their realization, as their availability will enable access to a wider set of tasks than those currently achievable.

Early experimental realizations of $W$ states relied on the blockade effect in Rydberg atom platforms to implement the creation of a single collective excitation, leading to $W$-like states in small atomic ensembles~\cite{Saffman2010, Ebert2015}. 
However, these approaches are inherently limited by the probabilistic nature of single excitation generation and do not readily scale to larger system sizes with high fidelity. More generally, while significant progress has been achieved in preparing multipartite entangled states across different platforms, including trapped ions, photonic systems, and superconducting processors~\cite{roos2004, haffner2005, kim2010, Eibl2004, grafe2014, Cruz2019}, the deterministic generation of large-scale $W$ states remains challenging. Indeed, in contrast to GHZ states, which can be prepared as ground states of suitable local Hamiltonians~\cite{Omran_2019}, $W$ states cannot arise as unique ground states of any local, gapped Hamiltonian~\cite{Gioia2023}. This fundamental obstruction prevents a straightforward extension of ground-state preparation protocols to $W$ states. 

Alternative strategies, such as measurement-based schemes that convert short-range into long-range entanglement, can in principle achieve high fidelity on digital platforms, but at the cost of probabilistic success rates that decrease with increasing fidelity~\cite{Buhrman2024, Piroli2024, farrell2025}.

\begin{figure}[t!]
    \centering
    \includegraphics[width=.5\textwidth]{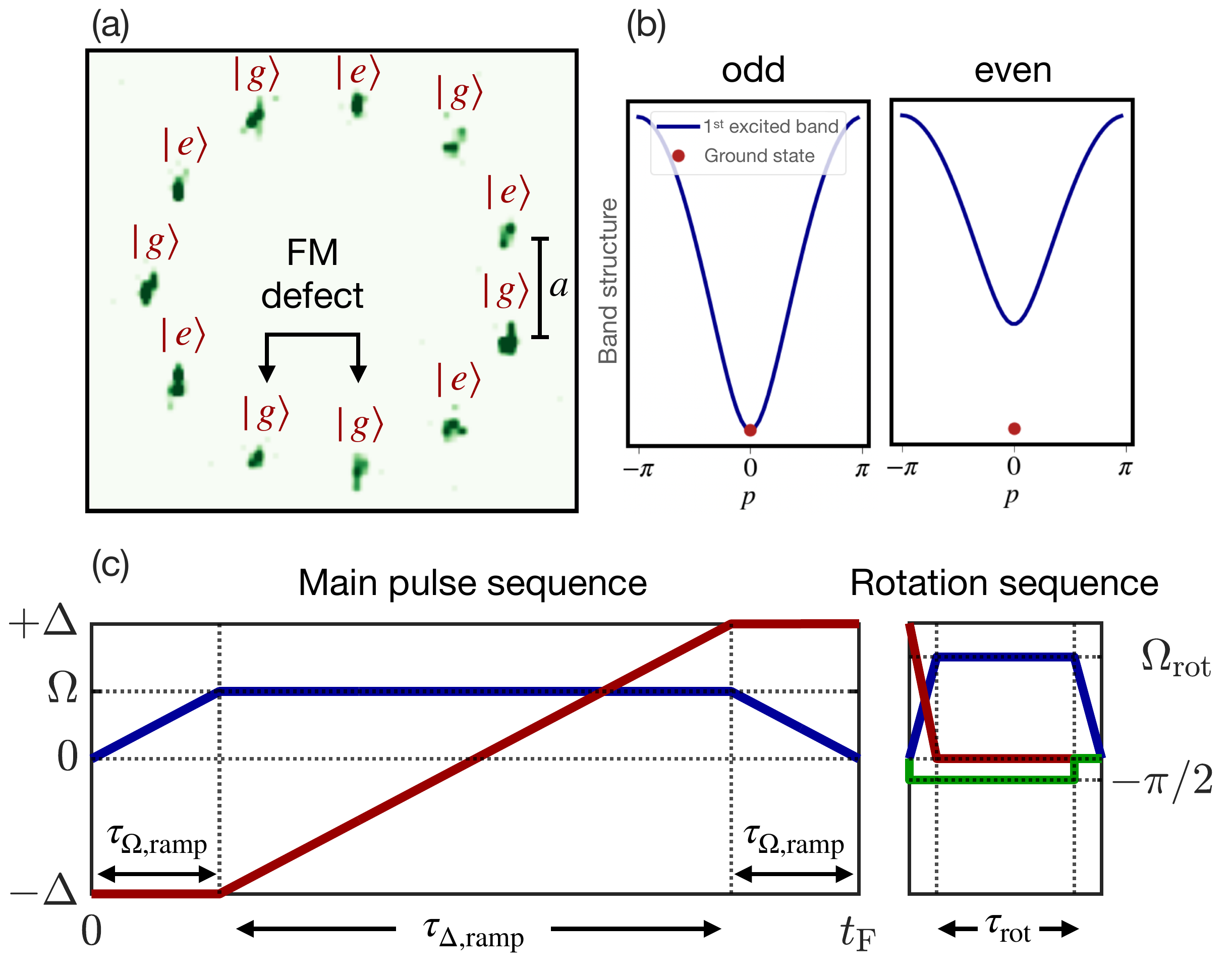}
    \caption{\textbf{Experimental setup to generate $W$ state through frustration and expose the generated many-body entanglement} \textbf{(a)} The setup geometry image taken by the Rydberg atom array Aquila~\cite{Aquila} where bright dots are atoms in their ground states forming a ring geometry with odd parity system size. Lattice distance is marked by $a$, and it is set to $a=7.1\mu$m in this specific image of $L=11$ atoms. An anti-ferromagnetic kink state is marked in red, where we point to the position of the ferromagnetic defect (see text). This is one viable experimental shot result after the main pulse sequence in (c) is applied. \textbf{(b)} The many-body energy gap decreases with system size as $L^{-2}$ in odd-parity chain sizes, leading to a gapless spectrum in the thermodynamic limit, whereas for even-parity chain sizes, the spectrum is gapped. \textbf{(c)} Experimental pulse sequence to prepare a many-body $W$ state where ramp times $\tau_{\Omega,\rm ramp}$, $\tau_{\Delta,\rm ramp}$, Rabi frequency $\Omega$ and detuning $\Delta$ are optimized for best $W$ state populations. The rotation sequence is applied in the second set of experiments to measure the state preparation fidelity by introducing a Rabi frequency phase, which changes the basis, green line, while setting the detuning to zero. Here, too, the evolution time $\tau_{\rm rot}$ and rotation Rabi frequency $\Omega_{\rm rot}$ are optimized.}
    \label{fig:fig0}
\end{figure}
%TC:endignore
These limitations highlight the need for a scalable and deterministic protocol to generate $W$ states with high fidelity.
Motivated by these considerations, we propose generating $W$ states in Rydberg atom arrays by arranging an odd number of atoms in a ring with effective antiferromagnetic (AFM) nearest-neighbor interactions. In this geometry, the system cannot satisfy all pairwise interactions simultaneously: one \textit{ferromagnetic} defect is necessarily introduced, which can occupy any position along the ring. Quantum mechanically, this defect delocalizes over all sites, giving rise to a coherent superposition that directly maps onto a $W$ state (see Fig.~\ref{fig:fig0}(a)). This mechanism is an instance of {\it topological ring frustration}~\cite{Dong2016, Maric2020_destroy, Maric2020_induced, Catalano2024_battery}, arising from the interplay between geometrical constraints (periodic boundary conditions with an odd number of sites) and antiferromagnetic interactions. In the classical limit of purely commuting interactions, frustration leads to an extensively degenerate ground-state manifold~\cite{Toulouse1986}, which is lifted by quantum fluctuations, producing a gapless low-energy spectrum in the thermodynamic limit. This behavior contrasts with even-length chains, where the spectrum remains gapped (Fig.~\ref{fig:fig0}(b))~\cite{Campostrini2015b, Dong2016, Maric2020_destroy, Maric2020_induced}. Crucially, this setting enables the adiabatic preparation of finite-size $W$ states without violating the no-go theorem of Ref.~\cite{Gioia2023}, although the rate of adiabatic cooling must be progressively reduced as the system size grows, in order to maintain high fidelity.

While the connection between $W$ states and frustrated antiferromagnetic spin systems is already apparent in the minimal case of a triangular loop, which has been used to generate three-qubit $W$ states in ion-trap experiments~\cite{kim2010}, this approach has not been extended to larger systems, where frustration could enable genuinely many-body state engineering. To address this challenge, we employ the {\it Aquila} Rydberg atom analog quantum simulator~\cite{Aquila}, developed by QuEra Computing. Its programmable geometry, enabled by movable optical tweezers, allows us to realize uniform ring configurations with an odd number of atoms (Fig.~\ref{fig:fig0}(a)), a key requirement for inducing topological frustration while minimizing disorder effects that could otherwise localize the defect.

As a final step, to validate our protocol, we address a central challenge in the field of quantum state preparation, that is the certification of many-body quantum states. Indeed, as full state tomography becomes infeasible for large system sizes, we introduce a Bayesian state-tomography scheme that leverages the close agreement between classical simulations and experimental data. Conceptually, our approach is related to the {\it classical shadow} framework~\cite{Huang2020, Elben2023, CIESLINSKI20241}, which employs randomized measurements for efficient state reconstruction. In contrast, we exploit deterministic Hamiltonian evolution combined with a small number of stochastic parameters to model experimental noise, significantly reducing the computational overhead. Let us remark that this method enables the certification of generic quantum states with tens of qubits, and provides robust lower bounds on their preparation fidelity.

\section{\texorpdfstring{$W$}{W}-State preparation}
The Hamiltonian of a Rydberg atom array is written as ($\hbar=1$) \cite{labuhn2016,bernien2017}
\begin{equation}
    \label{eq:Hryd}
    \hat H \!=\! \hat H_{\rm int} +  \sum_{\ell=1}^L \left(\frac{\Omega(t)}{2} e^{i \phi(t)}\ket{g_\ell}\bra{r_\ell} + \textrm{h.c.} -\Delta(t) \hat n_\ell \!\right)\!,\!
\end{equation}
where atoms are treated as two-level systems, i.e., Rydberg $\ket{r_l}$ and ground $\ket{g_l}$ states, and the interaction Hamiltonian $ \hat H_{\rm int} = \sum_{\ell<j}^L U_{\ell j}\hat n_\ell \hat n_j$ is governed by the van der Waals interactions of strength $U_{\ell j}=C_6/r_{\ell j}^6$. Here $r_{\ell j}$ is the distance between atoms at sites $\ell$ and $j$, $C_6=5.42\times 10^{-24} \textrm{ Hz m}^6$ is the Rydberg interaction coefficient for the Rydberg state $\ket{70 S_{1/2}}$ and $\hat n_\ell=\ket{r_\ell}\bra{r_\ell}$ counts the Rydberg excitations. The transitions of the atoms between Rydberg and ground states are captured by the amplitude $\Omega(t)$ and the phase $\phi(t)$ of the Rabi drive, while the detuning $\Delta(t)$ sets how off-resonant the Rabi drive is. The technical details and limitations of our experimental setup can be found in the {\it Aquila} whitepaper \cite{Aquila}.

%\st{To highlight the connection to the physics of frustrated magnetic rings,} 
This Hamiltonian can also be understood in terms of spin-$1/2$ degrees of freedom through the mapping $\ket{r\,(g)}\to\ket{\uparrow(\downarrow)}$. In operator terms, this corresponds to $\hat{n}_\ell=(1+\hat{\sigma}^z_\ell)/2$ and $\ket{g_l}\bra{r_l}=\hat{\sigma}^{-}_\ell$, in such a way that the Hamiltonian~\eqref{eq:Hryd}, in spin language, corresponds to a long-range Ising chain with a longitudinal and transverse field \cite{labuhn2016,PhysRevLett.120.180502,PhysRevA.110.053321,dag2024emergentdisordersubballisticdynamics}.
The effective AFM interaction is achieved when $a^6 \Omega / C_6 \ll 1$, $a$ being the lattice spacing, corresponding to the blockade regime in which two close atoms cannot both be excited to their Rydberg levels~\cite{Gaetan2009}. In this regime, and in the limit $\Omega \to 0$ (i.e., vanishing transverse field), when the total number of atoms $L$ is odd, the ground state of Eq.~\eqref{eq:Hryd} can be written as a superposition of AFM kink states with equal probabilities,
\begin{equation} \label{eq:GS}
    \ket{K_S}=\frac{1}{\sqrt{L}}\sum_{k=1}^L\ket{k}_{\rm AFM},
\end{equation}
where $\ket{k}_{\rm AFM}$ represents a Ne\'el state with a single ferromagnetic defect between sites $k$ and $k+1$, i.e., \mbox{$\ket{k}_{\rm AFM}=\ket{\dots\, {\color{red}r}\,{\color{blue}g}\,{\color{red}r}\, \stackrel{k}{{\color{blue}g}}\, \stackrel{}{{\color{blue}g}}\, {\color{red}r}\,{\color{blue}g}\,{\color{red}r}\, \dots}$.} One such kink state is depicted in Figure~\ref{fig:fig0}(a). We should note here that $\ket{K_S}$ is not exactly a $W$ state. While  $\ket{K_S}$ is not strictly identical to a W state at finite system size, it is operationally equivalent for quantum information tasks, as it can be efficiently mapped onto a W state via shallow Clifford circuits or with the aid of an ancilla qubit, without affecting its multipartite entanglement structure.
It is important to remark that most algorithms can be adjusted to accommodate this difference without impacting their performance. Moreover, the two states can be mapped to each other, either
through Clifford circuits~\cite{Odavic2023, Catalano2024_magic} or by employing an ancilla qubit (see Supplementary Material (SM)).
%TC:ignore
\begin{figure}[t]
    \centering
    \includegraphics[width=.4\textwidth]{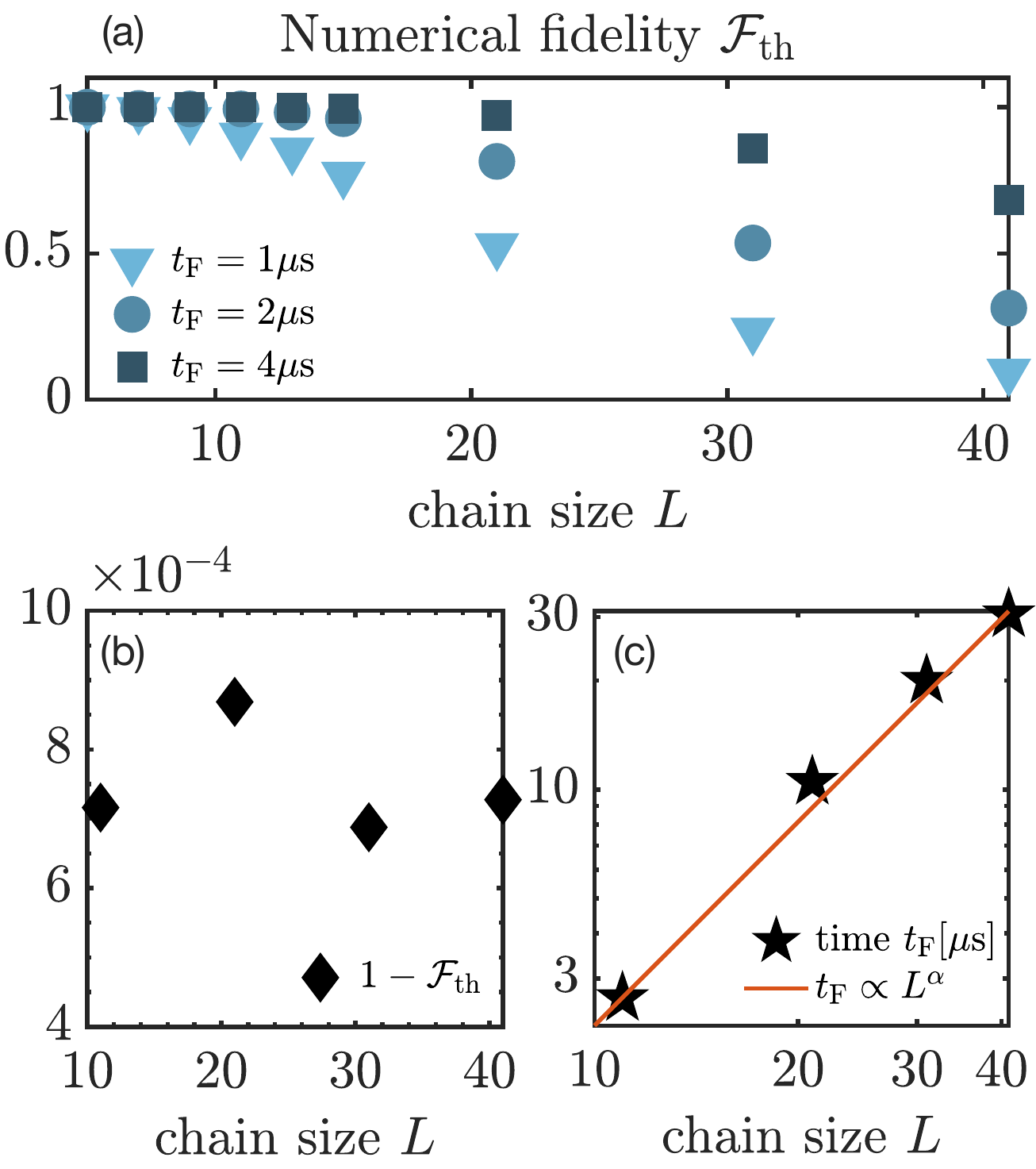}
    \caption{\textbf{Numerical simulation of the $W$ state preparation protocol} \textbf{(a)} The numerical fidelity of state preparation $\mathcal{F}_{\rm th}$ as a function of the system size, for atom spacing $a=6$ $\mu$m, Rabi frequency $\Omega=15$ rad$/\mu$s and total evolution time of $t_{\rm F}=1, 2, 4$  $\mu$s.  \textbf{(b)} Infidelity (that it, $1-\mathcal{F}_{\rm th}$), of the state preparation for slower protocols, and \textbf{(c)} corresponding time needed to achieve an infidelity smaller than $10^{-3}$ as a function of the system size $L$. These numerical data have been obtained for $\Omega=15$ rad$/\mu$s and $a=6$ $\mu$m.}
    \label{fig:fig2}
\end{figure}
%TC:endignore
To prepare $\ket{K_S}$, we initialized a ring of $L$ atoms (with $L$ odd) in their ground state, which is the ferromagnetic, separable ground state of Eq.~\eqref{eq:Hryd} for $\Omega = 0$ and large negative $\Delta$. We then implemented an adiabatic protocol consisting of alternating linear ramps of the detuning $\Delta (t)$ and Rabi frequency $\Omega (t)$ (Figure~\ref{fig:fig0}(c)), steering the system towards the target state. 
%While this protocol could be further optimized, our numerical simulations already show satisfactory results for system sizes much greater than those currently achievable on {\it Aquila}, which is limited by coherence times and grid size~\cite{Aquila}.

Denoting the instantaneous ground state during the evolution as $\ket{\psi(t)}$, we evaluate the fidelity with $\ket{K_S}$ at the end of the protocol as $\mathcal{F}_{\rm th}=\vert\braket{K_S \vert \psi(t_{\rm F}) }\vert^2$.
Figure~\ref{fig:fig2} reports the theoretical $\mathcal{F}_{\rm th}$ for various ring sizes $L$ at fixed lattice spacing $a=6\mu$m, Rabi frequency $\Omega=15$ rad/$\mu$s, and evolution time $t_{\rm F}$.
For each $L$, the fidelities were further optimized over detunings $\Delta\in[10,50]$ rad/$\mu$s (see SM). We used exact diagonalization up to $L=15$, while larger systems ($L=21,31,41$) were simulated with tensor-network techniques based on matrix product states (MPS), e.g., time-evolving block decimation~\cite{Catalano2024_TEBD}. These results assume ideal conditions: perfect ring geometry, unitary evolution, and uniform control fields. Later, we will incorporate dephasing, atom motion, and fluctuations in control fields to model experimental imperfections.

At fixed $t_{\rm F}$, the fidelity decreases as a function of the system size, Figure~\ref{fig:fig2}(a), reflecting the closing of the energy gap under frustration, see Figure~\ref{fig:fig0}(b). Consequently, without additional improvement of the protocol, longer adiabatic evolution times are required to maintain high fidelity as $L$ grows~\cite{Gioia2023}. In particular, if we further optimize over $t_{\rm F}$, we can retrieve infidelity smaller than $10^{-3}$. Results up to $L=41$ atoms are shown in Figure~\ref{fig:fig2}(b-c), highlighting that the optimal $t_{F}$ scales polynomially with the system size and exhibits an exponent of $\alpha=1.87\pm0.11$, consistent with the scaling of the smallest gap ($L^{-2}$) in a frustrated ring.

\section{Bayesian State validation protocol}

After state preparation, projective measurements along the $z$-axis (the axis of interaction) can probe the bit-string probabilities of $\ket{k}_{\rm AFM}$ states and detect unwanted strings. However, to guarantee that the prepared state is not just an incoherent mixture of $\ket{k}_{\rm AFM}$ states, one must also access the off-diagonal elements of the density matrix. Direct state tomography typically involves a number of measurements that grows exponentially with the number of qubits~\cite{Haah2016,Talath2025}. Moreover, even for small systems of just a few qubits, the impossibility to switch off Rydberg interactions during the experiment in the analog atom array produces non-trivial evolutions during any rotation sequence needed to change the basis of measurement~\cite{Omran_2019}. Therefore, full state tomography is not only cursed by exponential scaling with system's size, but it is also practically unfeasible in Rydberg atom array based analog simulators. 

Because of these fundamental limitations, we introduce a fundamentally different Bayesian paradigm for quantum state certification, in which prior physical knowledge of the system is incorporated directly into the reconstruction procedure. Rather than attempting a full reconstruction of the quantum state, our approach leverages on numerical simulations to generate a physically motivated ensemble of candidate states, whose statistical weights are refined through experimental observations.

The core of the Bayesian approach is the update formula
\begin{equation}
	P(H|E) = \frac{P(E|H) \cdot P(H)}{P(E)} \; ,
	\label{Bayes}
\end{equation}
where $P(H)$ is the {\it prior probability} reflecting the initial hypothesis $H$, and $E$ is the evidence, i.e., new data used to improve the estimate. $P(H|E)$ is the {\it posterior probability}, the probability of $H$ given $E$, which is the quantity of interest: the probability of a hypothesis given the observed evidence. $P(E|H)$ is the {\it likelihood}, which is the probability of observing $E$ given $H$. It indicates the compatibility of the evidence with the given hypothesis. $P(E)$ is the {\it marginal likelihood} and constitutes the normalization factor. Using this formula, one can improve the initial hypothesis through new data, yielding a more accurate probability distribution~\cite{MacKay2003}.

Our approach differs from the usual Bayesian quantum state tomography of~\cite{Schack2001, Blume2010, Granade2016, Lukens2020} in multiple ways. 
First, instead of starting from a null prior and progressively updating it through new measurements, we refine a prior generated by classical simulations using experimental data. 

Importantly, the numerical simulations are not assumed to provide an exact description of the experiment. 
Rather, they serve only to generate a physically motivated ensemble of candidate states, whose statistical weights are subsequently determined by their agreement with experimentally measured observables.
Crucially, our approach shifts the complexity of quantum state certification from the exponential scaling with the Hilbert space dimension to a scaling governed by a small number of physically motivated parameters describing the experimental setup.
%TC:ignore
 \begin{figure}[t]
    \centering
\includegraphics[width=.85\columnwidth]{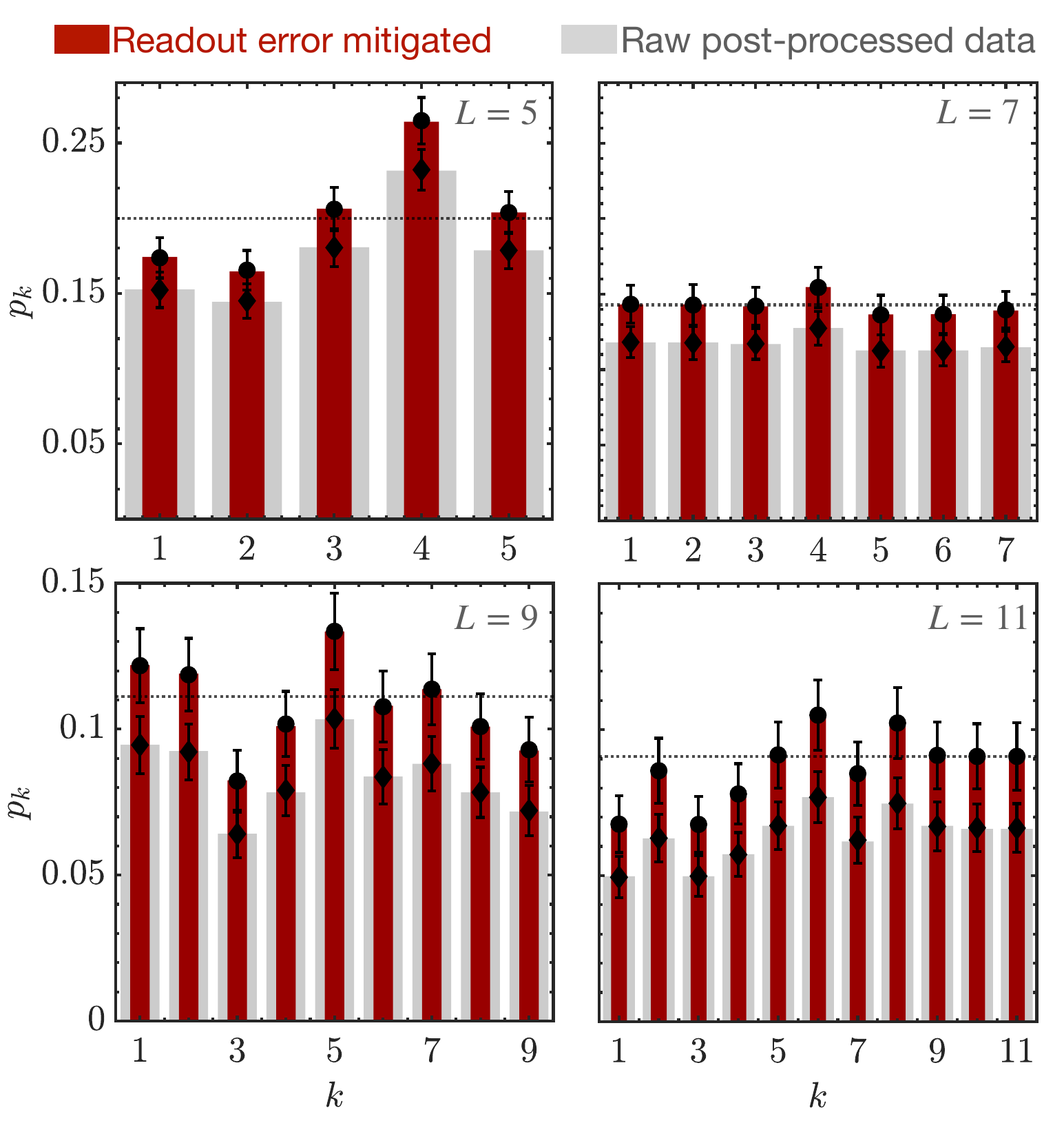}
    \caption{\textbf{Experimental results of $W$ state preparation protocol with chain sizes $L=5-11$.} The populations of the prepared state, where gray and red boxes are the raw and error-mitigated data \cite{Bravy2021}, see Methods for information on utilized error mitigation. The black error bars are computed via bootstrapping the experimental bit strings \cite{efron1979, efron1994}.
    The dashed-black horizontal lines mark $1/L$, which is the population value for a uniform $W$ state. For each experiment, 1000 bit-strings are collected, and more than $900$ of them are used in data analysis after post-processing based on the correct initial state. See SM for details.}
    \label{fig:fig3}
\end{figure}
%TC:endignore
In practice, we proceed as follows (details in Sec.~S3 of SM):
\begin{itemize}
    \item Numerically generate a distribution of states after the $W$ state preparation sequence, incorporating experimental uncertainties such as atom motion from thermal fluctuations, decoherence, and readout errors;
     \item Compare the experimentally observed and numerically generated bit-string frequencies, to select the optimal parameters and establish the prior distribution;
    \item Use the agreement (likelihood) between experimentally measured and numerically predicted bit-string frequencies after rotation to construct a posterior distribution. %in the $x$-basis and evaluate $P_{2n,k}^x$.
\end{itemize}

%\section{Experimental results and their Bayesian interpretation}
\section{Implementation on the Rydberg atom array platform}

Using a neutral atom simulator with an approximately square plate having an edge length of $\sim 75\mu$m~\cite{Aquila}, we are limited to rings of $L\leq11$ atoms to ensure uniformity of the ring avoiding laser noise. We study four ring sizes, $L=5$–$11$, optimizing lattice spacing and pulse sequences to prepare $W$ states with the highest fidelity achievable.

Figure~\ref{fig:fig3} shows the measured $z$-basis populations, i.e., the probabilities $p_k$ of $\ket{k}{\rm AFM}$ states, indicating that each kink state appears with roughly equal probability, as expected for a $W$ state. Residual non-uniformity among $\ket{k}_{\rm AFM}$ states arises from spatial inhomogeneities in detuning and Rabi frequencies, and from atom motion~\cite{dag2024emergentdisordersubballisticdynamics,Marcuzzi2017}, as confirmed by numerical simulations. Raw data (grey) include only post-processing based on the correct initial state, and a small fraction of non-$\ket{k}{\rm AFM}$ bit-strings is present due to such experimental imperfections and readout errors. The (readout) error-mitigated data (crimson) additionally correct for the readout errors (see Sec.~S5 in SM), yielding a probability of populating excited states below 0.05 for all chains. 

\begin{figure}
\centering
\includegraphics[width=0.85\columnwidth]{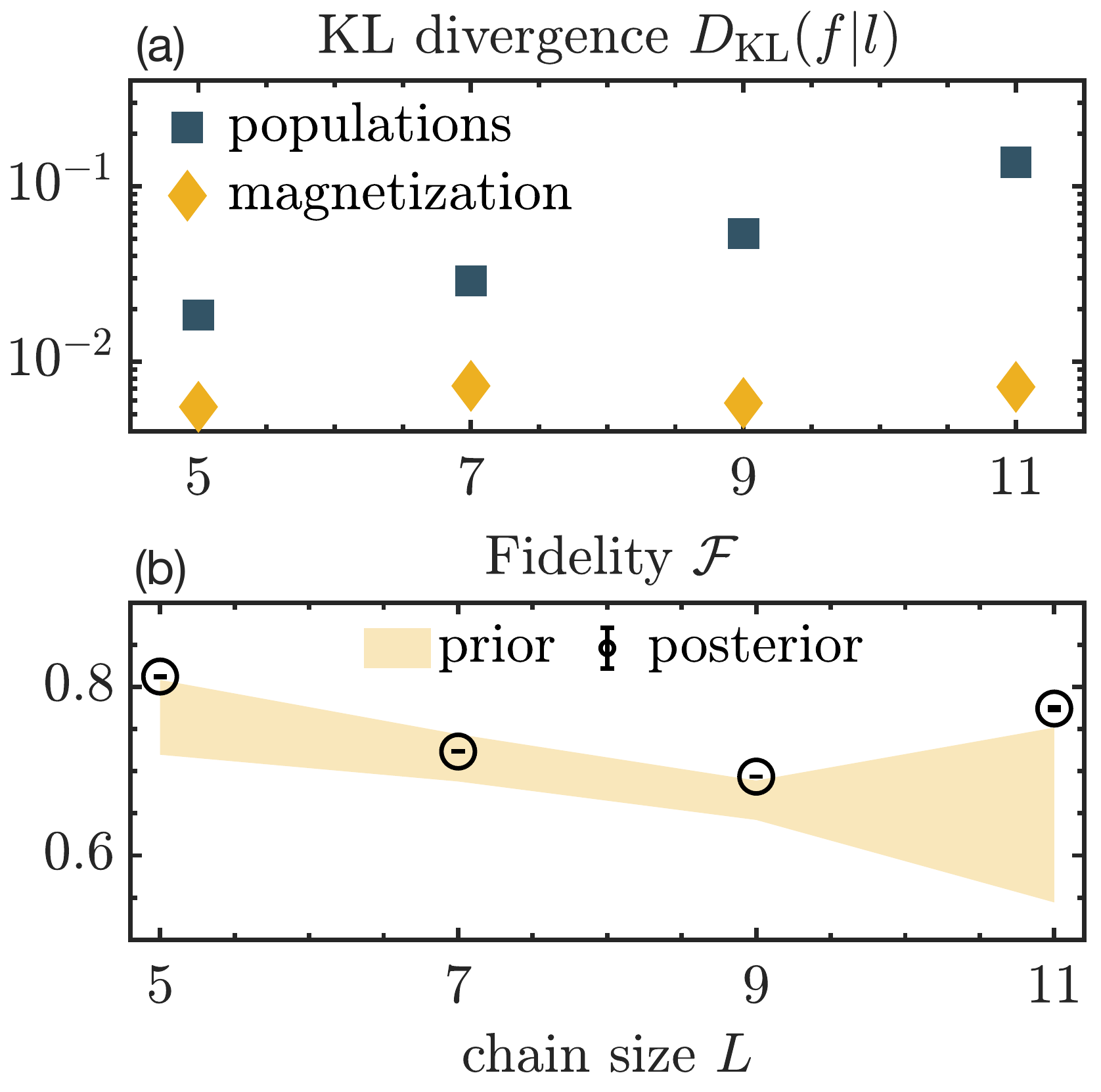}
\caption{\textbf{Measured $W$ state fidelity conditioned on the theoretical simulations} \textbf{(a)} Kullback-Leibler divergence between experimental and numerical distributions. The blue squares refer to the KL divergence between the bit-string distributions, i.e.,~the outcomes of the projective measurement on the quantum state after the main pulse sequence. The yellow diamonds represent the KL divergence between the distributions of total magnetization, i.e., the frequencies of states with a given magnetization, regardless of the microscopic bit-string distribution. Since a perfect state preparation would yield always the same magnetization, the agreement between experimental and numerical distributions is an excellent indicator of the accuracy of our numerical modeling.
\textbf{(b)} The fidelity of the prepared states: the shaded area correspond to the fidelity spread according to the numerical prior distribution, which is created using just the diagonal ($z$-basis) experimental data. The markers are the result of the Bayesian procedure in eq.~\eqref{eq:BayesianUpdate} and are computed with the posterior probabilities which take into account the results of the (off-diagonal) measurements after rotation. The generated states are found to be quantum correlated as the fidelity is much larger than $1/L$ for all system sizes.}
\label{fig:fig4}
\end{figure}

Figure~\ref{fig:fig4}(a) compares experimental and numerical bit-string frequencies $p_k$ using the Kullback–Leibler divergence $D_{\rm KL}(f||t)$~\cite{Cover2006}, which quantifies the agreement between the measured ($f$) and numerical ($t$) distribution. For a set of bit-strings ${\vec{x}}$, this is defined as
\begin{equation}
    D_{\rm KL} (f||t) \equiv \sum_{\{\vec{x}\}} 
	 \left( f (\vec{x}) \log f (\vec{x}) - f (\vec{x}) \log t (\vec{x}) \right) ,
    \label{KLDef}
\end{equation}
and is minimized (vanishing) when the two distributions coincide. 
Figure~\ref{fig:fig4}(a) shows that for a larger number of atoms the $D_{\rm KL}$ increases (but remains small), likely due to statistical fluctuations, induced by the fact that the number of samples taken does not grow sufficiently to properly characterize the bit-string distribution. %exponentially in $L$ as the Hilbert space. 
This interpretation is corroborated by the magnetization in Figure~\ref{fig:fig4}(a). For a perfect $W$ state, there is no quantum uncertainty in magnetization and all measurements should provide the same result. The excellent, size-independent agreement between experiment and simulation indicates how accurately our numerical model captures stochastic experimental uncertainties.
%TC:ignore
%TC:endignore

In Figure~\ref{fig:fig4}(b) we show the fidelity relative to a perfect $W$ state. The shaded region represents the spread of fidelities, measured using the prior distribution selected after comparison with experimental populations. The markers represent the Bayesian-updated fidelities, which additionally incorporate measurements after the rotation sequence.
Data from ${\cal M}=5$ rotation experiments, each producing ${\cal N}^{(\alpha)}=1000$–$3000$ samples, are used to construct the likelihoods ${\cal L}^{(\alpha)}_j \equiv \exp \left[- {\cal N}^{(\alpha)} D_{\rm KL} (f^{(\alpha)}||t^{(\alpha)} (\rho_j)) \right]$. In turn, these are used to compute new statistical weights which take into account the agreement between numerical simulations and experimental results, i.e.,
\begin{equation}
w_j=\frac{\prod_{\alpha=1}^5\mathcal{L}_j^{(\alpha)}}{\sum_{\ell=1}^{\mathcal{Q}}\prod_{\alpha=1}^5\mathcal{L}_\ell^{(\alpha)}},
\end{equation}
%Applying Bayes' formula (Eq.~\eqref{eq:Bayes}), the posterior probabilities for output bit-strings are
%\begin{eqnarray}    \label{eq:BayesianUpdate}    p_B (\vec{x}) & \equiv & \frac{ \sum_{j=1}^{\cal Q} \left( \prod_{\alpha=1}^{{\cal M}} {\cal L}^{(\alpha)}_j \right) p_j(\vec{x}) }{ {\cal Z} } \; , \\    {\cal Z} & \equiv & \sum_\mu \sum_{j=1}^{\cal Q}  \left( \prod_{\alpha=1}^{{\cal M}-1} {\cal L}^{(\alpha)}_j \right)  p_j(\vec{x}) \; ,\end{eqnarray}
%starting from prior probabilities $p_j(\vec{x})$ derived from 
derived for ${\cal Q}=100$ numerically generated density matrices $\hat{\rho}_j$ (see SM). Then, the inferred fidelity is given by
\begin{equation}  
\label{eq:BayesianUpdate}
\mathcal{F}=\sum_{j=1}^\mathcal{Q} w_j \mathcal{F}_{e,j},
\end{equation}
where $\mathcal{F}_{e,j}$ is the fidelity computed from the $j$-th simulation.  We observe in Figure~\ref{fig:fig4}(b) that for $L=11$ the (prior) numerical expectation value for the fidelity has a large spread (consistently with the larger $D_{\rm KL}$ observed in Figure~\ref{fig:fig4}(a)), but the Bayesian update reduces this uncertainty and brings the fidelity in line with the other data. It should also be noted that the experiment with $L=11$ atoms has been performed with a larger lattice constant, $a=7.1\mu$m, compared to other rings with $a=6\mu$m to accommodate the larger ring on the plate. This might explain the difference from the rest of the chain sizes, as the increased lattice constant reduces the strength of Rydberg interactions and enhances the timescale of decoherence (see sec.~S3.B of the Supplementary Material). 

Let us also note that the fidelities obtained in the experiment are lower than what is predicted in ideal simulations and reported in Fig.~2, mainly due to experimental imperfections such as laser noise, atom motion and decoherence as well as the additional effects introduced by in the state validation, which requires  the additional the rotation sequence after the state is prepared. 

\section{Conclusions and outlook}

We have proposed and experimentally implemented a protocol for the adiabatic preparation of $W$ states, a central class of multipartite entangled states with broad applications in quantum information science. 
Our approach exploits topological frustration in a ring geometry with an odd number of atoms, where the interplay of boundary conditions, antiferromagnetic interactions, and quantum interactions gives rise to a delocalized excitation corresponding to a $W$ state. 
This is realized on a programmable neutral-atom quantum simulator, demonstrating a scalable and deterministic route to preparing $W$ states in many-body systems.
We remark that, although the protocol naturally targets odd system sizes, even-size $W$ states can also be obtained via partial tracing, as discussed in the Supplemental Material. This stems from the resilient structure of $W$ states, which, unlike GHZ, remain entangled even after partial measurements.

A central challenge in the experimental realization of highly entangled many-body states is their certification, as full state tomography becomes infeasible at increasing system sizes. To address this problem, we developed a new Bayesian tomography scheme that combines accurate numerical simulations with experimental data obtained after controlled dynamical evolution, enabling the inference of otherwise inaccessible observables. 

Using this approach, we estimate a fidelity of $\mathcal{F} = 0.774$ for the largest system of $L = 11$ atoms on {\it Aquila}. 
The reported fidelities should be interpreted as conservative estimates, as the validation protocol introduces additional imperfections—primarily due to the rotation sequence—that are not present in the prepared state.
Importantly, the Bayesian method introduced here is not specific to W states, establishing a general and scalable framework for quantum state certification. This is particularly significant in analog quantum simulators based on Rydberg atom arrays, where one cannot perfectly rotate the measurement basis because of unavoidable interaction effects. By leveraging physically informed priors, it enables the extraction of global properties of many-body quantum states with a cost that scales with the number of relevant experimental parameters rather than the size of the Hilbert space. 
Crucially, compared to similar approaches, it is efficient and scales with the number of implemented stochastic parameters, and not of qubits. 

The main limitation in scaling the state preparation to larger system sizes is set by experimental constraints on atom spacing and coherence, which currently restrict the maximum achievable system size. These limitations are not fundamental, depend on the experimental platform, and can be alleviated by improved control over lattice geometry and noise reduction in Rydberg platforms. For instance, atom distances of 1$\mu$m and 0.45$\mu m$ have already been realized, which would accommodate $L < 18$ and $L < 40$ atoms respectively. Looking ahead, several strategies can further enhance scalability. 
First, shortcut-to-adiabaticity techniques~\cite{Lukin2024} could reduce the preparation time while preserving high fidelity. Second, optimal control methods~\cite{Doria2011, Lloyd2014} may enable the design of more efficient pulse sequences beyond simple linear ramps. Finally, implementing the protocol in two-level systems encoded in hyperfine states of neutral atoms~\cite{bluvstein2022} offers significantly longer coherence times, providing a promising route toward larger-scale realizations. Moreover, %let us highlight that, 
since our protocol is based on the adiabatic preparation of a Hamiltonian ground state, it %our protocol 
can easily be extended to different devices beyond Rydberg quantum simulators, such as trapped ion platforms or digital machines that can simulate trotterized time evolution.

As the rapid progress of quantum technologies calls for scalable methods to prepare highly entangled states with complex correlation structures, $W$ states and, more generally, Dicke states represent key resources for quantum information processing. While gate-based approaches can generate such states, they often rely on probabilistic protocols, which limits their scalability. Our results establish topological ring frustration as a powerful mechanism for the deterministic preparation of multipartite entangled states in analog quantum simulators. Indeed, considering long-range or two-dimensional extension of topologically frustrated Hamiltonians, addressable on trapped ions or Rydberg atom arrays as well as on digital machines, one could engineer coherent superpositions of states with multiple excitations~\cite{Torre2025,Catalano2025_2D} and open the doors to the realization of more complex states, characterized by unique quantum resources. 

To conclude, %let us highlight that, 
combined with recent advances in Rydberg atom arrays, including the ability to transport entanglement via controlled motion of tweezers that hold atoms~\cite{bluvstein2022}, our preparation scheme leads to new opportunities for implementing quantum communication protocols such as secure multiparty communication~\cite{Wang2007, chen2008} and quantum teleportation~\cite{Joo2003}. More broadly, hybrid analog–digital architectures based on neutral atoms provide a promising route toward scalable quantum processors, where entangled resource states prepared via analog protocols can be harnessed for practical quantum algorithms and technologies~\cite{GeimArxiv2026}.

\section{Methods}
\subsection{Numerical simulations}
We performed our numerical analysis with the aid of both exact diagonalization and tensor network based algorithms. In particular, in order to simulate the ideal (unitary) dynamics under the action of the Rydberg hamiltonian Eq.~\eqref{eq:Hryd}, we used exact diagonalization to simulate system's of sizes up to $L=15$. For larger system's sizes we considered tensor networks techniques, and encoded the physical states using matrix product states (MPS) ansatze. In this case, time evolution was performed using a time evolution block decimation (TEBD) algorithm, based on the matrix product operator (MPO) method introduced in on~\cite{Catalano2024_TEBD} for the representation of the trotterized time evolution operator. In order to achieve convergence, we used a maximum bond dimension of $\chi=64$ for $L=41$. 

\begin{table*}[t]
\centering
\caption{The parameter values used in experiments. Total $5$ different experiments run with $5$ different values of $\Omega_{\rm rot}$ for each chain size, all of which are used in data presented in Figure~\ref{fig:fig4}(b).}
    \label{tab:table1}
    \begin{center}
    \begin{tabular}{c|c|c|c|c|c|c|c} % <-- 
      Chain size $L$ & Lattice spacing a ($\mu$m) & $\Omega$ (rad$/\mu$s) & $\Delta$ (rad$/\mu$s) & $t_{\rm F}$ ($\mu$s) & $\tau_{\Omega,\rm ramp}$ ($\mu$s) & $\tau_{\rm rot}$ ($\mu$s) & $\Omega_{\rm rot}$ (rad$/\mu$s) \\
      \hline
      5 & 6 & 11.46  & 29 & 2 & 0.25 & 0.15 & 7.97 $\pm$ 0.5,1 \\
      7 & 6 & 11.46 & 26 & 2 & 0.25 & 0.15 & 8.47 $\pm$ 0.5,1 \\
      9 & 6 & 11.46 & 29 & 2 & 0.25 & 0.15 & 7.21 $\pm$ 0.5,1\\
      11 & 7.1 & 11.46 & 25 & 3 & 0.5 & 0.2 & 6.99 $\pm$ 0.5,1\\
      \hline
    \end{tabular}
    \end{center}
\end{table*}
In order to simulate the experimental conditions, we considered the Rydberg Hamiltonian up to next nearest neighbor interactions,
\begin{equation}
    \label{eq:Hryd_noisy}
    H/\hbar=\sum_{\ell<j}\frac{C_6}{r_{\ell j}}n_\ell n_j+\sum_{\ell=1}^L\frac{\Omega_\ell}{2}\sigma^x_\ell-\sum_{\ell=1}^L \Delta_\ell n_\ell.
\end{equation}
Experimentally measured global fluctuations in the Rabi frequency and detuning were used to mimic the shot-to-shot noise, see next section and the SM for more details. In order to take into account the effect of thermal atom motion, each simulation was initialized with a random lattice, in which the position of each atom was extracted from the Gaussian distributions centered around the sites of an ideal ring with lattice constant $a$ and with standard deviation of $\delta x=\delta y=0.15\mu m$ for each of the coordinates. We also included local dephasing in our simulations, modeling the time evolution of the system through the following Lindblad master equation,
\begin{equation}
    \label{eq:lindblad}
    \dot{\rho}=-i[H,\rho]+\frac{\gamma}{2}\sum_{\ell=1}^L(2n_\ell \rho n_\ell - \{\rho,n_l\}),
\end{equation}
These jump operators implement local Rydberg-occupation dephasing. Physically, this form is appropriate for noise channels that couple to the Rydberg number operator, such as laser phase or detuning noise, and it can also serve as an effective phenomenological description of decoherence induced by off-resonant intermediate-state scattering~\cite{fang2024probingcriticalphenomenaopen,PhysRevA.97.053803,kozlej2025adiabaticstatepreparationthermalization}. We find that modeling these decoherence processes yields results consistent with the experimental data. Since the experimental evolution times are much shorter than the device-level Rydberg population lifetime, we do not include an explicit Rydberg population relaxation channel in our simulations.

\begin{table}[t]
\centering
\caption{The sampling numbers for the experiments in Figure~\ref{fig:fig3}.}
    \label{tab:table2}
    \begin{center}
    \begin{tabular}{c|c} % <-- 
      Chain size $L$ & The number of used shots  \\
      \hline
      5 & 963 \\
      7 & 941 \\
      9 &  919 \\
      11 & 922 \\
      \hline
    \end{tabular}
    \end{center}
\end{table}
After the simulation of state preparation, to compare the simulated bit-strings distributions with the experimental ones, we also implemented readout errors, simulating the possibility of bit flips during the projective measurements. 
Comprehensive details on the parameters used in the simulations are reported in Sec.~S5 of the supplementary material. From a technical standpoint, here we also employed tensor network methods based on MPS and MPOs. In particular, we represented the vectorized density matrix of the system as an MPS, and the Liouvillian super-operator as an MPO~\cite{Collura2024}. We then employed the time dependent variational principle algorithm (TDVP) with single site update to time evolve the system~\cite{Paeckel2019}. The largest bond dimension used was $\chi=64$ for $L=11$. 

\subsection{Experimental methods}

Before conducting many-body experiments, we calibrate our setup using a two-photon resonance measurement on an ensemble of non-interacting atoms confined in a two-dimensional geometry. In this protocol, the Rabi frequency is abruptly switched from zero to a target value, $\Omega$, while holding the detuning fixed at $\Delta$. A measurement is then performed at $t_{\pi}=\pi/\Omega$. Repeating this procedure for a range of detunings yields the excitation probability as a function of $\Delta$. The theoretical form of this probability, obtained through standard quantum optics methods~\cite{Meystre2007}, is $P_e(\Delta) = \frac{\Omega^2}{\Omega^2+\Delta^2} \sin^2\left(\sqrt{\Omega^2+\Delta^2} \frac{t_{\pi}}{2}\right)$.

By fitting the experimental data to this analytic expression, we extract both the overall detuning offset and the calibrated value of the Rabi frequency. This calibration establishes a direct correspondence between the programmed drive strength and the effective measured value. All subsequent many-body measurements are corrected using this relation. (See SM for long-term stability data on Rabi frequency, which demonstrates that the calibration remains consistent over months.) We also observe that the resulting population histograms are highly robust against parameter variations such as Rabi frequency or lattice spacing, provided the ramp efficiently cools the array into its AFM ground state. See SM for additional experimental data on populations for different sets of parameters.

The calibration further provides an estimate of the statistical uncertainty. Using a bootstrap analysis~\cite{efron1979, efron1994}, we confirm that the distributions of the calibrated Rabi frequency and detuning converge to Gaussian form and extract the corresponding $1\sigma$ errors. These uncertainties are incorporated into numerical simulations to model shot-to-shot fluctuations. Finally, we mitigate readout errors in the many-body data by applying confusion matrix corrections \cite{Bravy2021}. See SM for further details. 

The specifications of the experiments for different system sizes are shared in Table~\ref{tab:table1}. The fast ramps in the rotation sequence are done in 50ns. Population experimental results shared in Figure~\ref{fig:fig3} are for rings positioned at the center of the plate. See SM for experimental results of the state preparation when the rings shifted $\pm 10\mu$m from the center of the plate.

In total, 1000 bit-strings were collected in each experiment. However, not every experiment starts at the correct initial state, or it was thought so due to readout error. The corresponding final bit-strings are discarded. See Table~\ref{tab:table2} for the exact number of shots used to obtain Figure~\ref{fig:fig3}. To arrive at the fidelity values reported in Figure~\ref{fig:fig4}(b) and the KL divergence in Figure~\ref{fig:fig4}(a), we used data from experiments with different sets of parameters such that the prior distribution was sufficiently improved. Specifically, these data include five experiments at each system size with different $\Omega_{\rm rot}$, see Table~\ref{tab:table1}, and at different position for the ring, e.g.,~whether the center of the ring is centered on the plate or shifted to $\pm 10\mu$m, where each experiment has a shot number between $900-1000$ after post-processing based on the initial state. For specific and detailed information, see SM.

\section{Acknowledgments}

Authors thank Fang~Fang, Vinko Zlati\'c, Tommaso Macri, Lorenzo Piroli, for stimulating and helpful discussions, Amazon Web Services for the Cloud Credit for Research Program and F100 Initiative start-up funds by Indiana University to run {\it Aquila} in this project, QuEra Computing for technical assistance on {\it Aquila}, Majd Hamdan for the machine image in Figure~\ref{fig:fig0}. This research was supported in part by grant NSF PHY-2309135 to the Kavli Institute for Theoretical Physics (KITP). C.B.D acknowledges support from the NSF through a grant for ITAMP at Harvard University. A.G.C. acknowledges support from the MOQS ITN programme, a European Union’s Horizon 2020 research and innovation program under the Marie Sk\l{}odowska-Curie grant agreement number 955479, and from the European Union’s Horizon Europe research and innovation program under the Marie
Sklodowska-Curie Grant No. 101206552 (QUEST). G.T. has been partially supported by the Slovenian Quantum Science Hub — SQUASH programme, co-funded by the European Union under the Horizon Europe Marie Skłodowska-Curie Actions COFUND programme, grant agreement No. 101177446, and by the Slovenian Research and Innovation Agency (ARIS), contract No. 5110-18/2025-5. SMG, and FF also acknowledge support from the project "Implementation of cutting-edge research and its application as part of the Scientific Center of Excellence for Quantum and Complex Systems, and Representations of Lie Algebras", Grant No. PK.1.1.10.0004, co-financed by the European Union through the European Regional Development Fund- Competitiveness and Cohesion Programme 2021-2027, from the Croatian Science Foundation (HrZZ) through the project IP-2025-02-1667, Mining the Quantum: Frustration, Disorder, and Devices.
The research leading to these results has received funding from the following organizations: Italian Research Center on HPC, Big Data and Quantum Computing (NextGenerationEU Project No. CN00000013), project EuRyQa (Horizon 2020); Italian Ministry of University and Research (MUR) via: Quantum Frontiers (the Departments of Excellence 2023-2027); the World Class Research Infrastructure - Quantum Computing and Simulation Center (QCSC) of Padova University; Istituto Nazionale di Fisica Nucleare (INFN): iniziativa specifica IS-QUANTUM. 

\section{Author contributions}
Every signing author has contributed to all aspects of this work and its writing. Every aspect of its development has been discussed collegially and ideas from everyone enriched the final outcome. In particular,  C.B.D. lead the experimental part of this work, performed the measurements, and the main data analysis. A.G.C. and G.T. coded and implemented the numerical simulations, designed the pulse sequences, and derived the estimator for the fidelity with respect to an ideal W state. F.F. had a leading role in the development of the Bayesian state tomography approach introduced in this work. S.M.G. contributed to the analysis of the numerical data and helped critically toward their interpretation, also with respect to the existing literature. All authors discussed the results, wrote parts of the original draft of the manuscript, and helped to bring it to its final form.
\section{Competing interests}
There are no competing interests to declare.
\section{Data and materials availability} The data related to this work are available in the following data repository: https://doi.org/10.7910/DVN/QGBEXF

\bibliographystyle{naturemag}
\bibliography{ref.bib} 

@article{Chitambar2019,
  title = {Quantum resource theories},
  author = {Chitambar, Eric and Gour, Gilad},
  journal = {Rev. Mod. Phys.},
  volume = {91},
  issue = {2},
  pages = {025001},
  numpages = {48},
  year = {2019},
  month = {Apr},
  publisher = {American Physical Society},
  doi = {10.1103/RevModPhys.91.025001},
  url = {https://link.aps.org/doi/10.1103/RevModPhys.91.025001}
}

@article{Horodecki2009,
  title = {Quantum entanglement},
  author = {Horodecki, Ryszard and Horodecki, Pawe\l{} and Horodecki, Micha\l{} and Horodecki, Karol},
  journal = {Rev. Mod. Phys.},
  volume = {81},
  issue = {2},
  pages = {865--942},
  numpages = {0},
  year = {2009},
  month = {Jun},
  publisher = {American Physical Society},
  doi = {10.1103/RevModPhys.81.865},
  url = {https://link.aps.org/doi/10.1103/RevModPhys.81.865}
}

@article{Plenio2007,
author = {Plbnio, Martin B. and Virmani, Shashank},
title = {An introduction to entanglement measures},
year = {2007},
issue_date = {January 2007},
publisher = {Rinton Press, Incorporated},
address = {Paramus, NJ},
volume = {7},
number = {1},
issn = {1533-7146},
journal = {Quantum Info. Comput.},
month = jan,
pages = {1–51},
numpages = {51}
}

@article{Ollivier2001,
  title = {Quantum Discord: A Measure of the Quantumness of Correlations},
  author = {Ollivier, Harold and Zurek, Wojciech H.},
  journal = {Phys. Rev. Lett.},
  volume = {88},
  issue = {1},
  pages = {017901},
  numpages = {4},
  year = {2001},
  month = {Dec},
  publisher = {American Physical Society},
  doi = {10.1103/PhysRevLett.88.017901},
  url = {https://link.aps.org/doi/10.1103/PhysRevLett.88.017901}
}

@article{Modi2012,
  title = {The classical-quantum boundary for correlations: Discord and related measures},
  author = {Modi, Kavan and Brodutch, Aharon and Cable, Hugo and Paterek, Tomasz and Vedral, Vlatko},
  journal = {Rev. Mod. Phys.},
  volume = {84},
  issue = {4},
  pages = {1655--1707},
  numpages = {0},
  year = {2012},
  month = {Nov},
  publisher = {American Physical Society},
  doi = {10.1103/RevModPhys.84.1655},
  url = {https://link.aps.org/doi/10.1103/RevModPhys.84.1655}
}

@article{Baumgratz2014,
  title = {Quantifying Coherence},
  author = {Baumgratz, T. and Cramer, M. and Plenio, M. B.},
  journal = {Phys. Rev. Lett.},
  volume = {113},
  issue = {14},
  pages = {140401},
  numpages = {5},
  year = {2014},
  month = {Sep},
  publisher = {American Physical Society},
  doi = {10.1103/PhysRevLett.113.140401},
  url = {https://link.aps.org/doi/10.1103/PhysRevLett.113.140401}
}

@article{Streltsov2017,
  title = {Colloquium: Quantum coherence as a resource},
  author = {Streltsov, Alexander and Adesso, Gerardo and Plenio, Martin B.},
  journal = {Rev. Mod. Phys.},
  volume = {89},
  issue = {4},
  pages = {041003},
  numpages = {34},
  year = {2017},
  month = {Oct},
  publisher = {American Physical Society},
  doi = {10.1103/RevModPhys.89.041003},
  url = {https://link.aps.org/doi/10.1103/RevModPhys.89.041003}
}

@article{Howard2017,
  title = {Application of a Resource Theory for Magic States to Fault-Tolerant Quantum Computing},
  author = {Howard, Mark and Campbell, Earl},
  journal = {Phys. Rev. Lett.},
  volume = {118},
  issue = {9},
  pages = {090501},
  numpages = {6},
  year = {2017},
  month = {Mar},
  publisher = {American Physical Society},
  doi = {10.1103/PhysRevLett.118.090501},
  url = {https://link.aps.org/doi/10.1103/PhysRevLett.118.090501}
}

@article{Leone2022,
  title = {Stabilizer R\'enyi Entropy},
  author = {Leone, Lorenzo and Oliviero, Salvatore F. E. and Hamma, Alioscia},
  journal = {Phys. Rev. Lett.},
  volume = {128},
  issue = {5},
  pages = {050402},
  numpages = {5},
  year = {2022},
  month = {Feb},
  publisher = {American Physical Society},
  doi = {10.1103/PhysRevLett.128.050402},
  url = {https://link.aps.org/doi/10.1103/PhysRevLett.128.050402}
}

@article{Dur2000,
   title={Three qubits can be entangled in two inequivalent ways},
   volume={62},
   ISSN={1094-1622},
   url={http://dx.doi.org/10.1103/PhysRevA.62.062314},
   DOI={10.1103/physreva.62.062314},
   number={6},
   journal={Physical Review A},
   publisher={American Physical Society (APS)},
   author={Dür, W. and Vidal, G. and Cirac, J. I.},
   year={2000},
   month=nov }

@article{Joo2003,
doi = {10.1088/1367-2630/5/1/136},
url = {https://dx.doi.org/10.1088/1367-2630/5/1/136},
year = {2003},
month = {oct},
publisher = {},
volume = {5},
number = {1},
pages = {136},
author = {Joo, Jaewoo and Park, Young-Jai and Oh, Sangchul and Kim, Jaewan},
title = {Quantum teleportation via a $W$ state},
journal = {New Journal of Physics}
}

@article{Wang2007,
doi = {10.1088/0253-6102/48/4/013},
url = {https://dx.doi.org/10.1088/0253-6102/48/4/013},
year = {2007},
month = {oct},
publisher = {},
volume = {48},
number = {4},
pages = {637},
author = {Wang Jian and Zhang Quan and Tang Chao-Jing},
title = {Quantum Secure Communication Scheme
with W State},
journal = {Communications in Theoretical Physics}
}

@article{Vijayan2020,
  doi = {10.22331/q-2020-08-03-303},
  url = {https://doi.org/10.22331/q-2020-08-03-303},
  title = {A robust {W}-state encoding for linear quantum optics},
  author = {Vijayan, Madhav Krishnan and Lund, Austin P. and Rohde, Peter P.},
  journal = {{Quantum}},
  issn = {2521-327X},
  publisher = {{Verein zur F{\"{o}}rderung des Open Access Publizierens in den Quantenwissenschaften}},
  volume = {4},
  pages = {303},
  month = aug,
  year = {2020}
}

@article{Grover1997,
  title = {Quantum Mechanics Helps in Searching for a Needle in a Haystack},
  author = {Grover, Lov K.},
  journal = {Phys. Rev. Lett.},
  volume = {79},
  issue = {2},
  pages = {325--328},
  numpages = {0},
  year = {1997},
  month = {Jul},
  publisher = {American Physical Society},
  doi = {10.1103/PhysRevLett.79.325},
  url = {https://link.aps.org/doi/10.1103/PhysRevLett.79.325}
}

@article{PhysRevLett.120.180502,
  title = {Detailed Balance of Thermalization Dynamics in Rydberg-Atom Quantum Simulators},
  author = {Kim, Hyosub and Park, YeJe and Kim, Kyungtae and Sim, H.-S. and Ahn, Jaewook},
  journal = {Phys. Rev. Lett.},
  volume = {120},
  issue = {18},
  pages = {180502},
  numpages = {5},
  year = {2018},
  month = {May},
  publisher = {American Physical Society},
  doi = {10.1103/PhysRevLett.120.180502},
  url = {https://link.aps.org/doi/10.1103/PhysRevLett.120.180502}
}

@article{PhysRevA.110.053321,
  title = {Motional decoherence in ultracold-Rydberg-atom quantum simulators of spin models},
  author = {Zhang, Zewen and Yuan, Ming and Sundar, Bhuvanesh and Hazzard, Kaden R. A.},
  journal = {Phys. Rev. A},
  volume = {110},
  issue = {5},
  pages = {053321},
  numpages = {9},
  year = {2024},
  month = {Nov},
  publisher = {American Physical Society},
  doi = {10.1103/PhysRevA.110.053321},
  url = {https://link.aps.org/doi/10.1103/PhysRevA.110.053321}
}

@article{chen2008,
  title = {CONTROLLED QUANTUM SECURE DIRECT COMMUNICATION WITH W STATE},
  volume = {06},
  ISSN = {1793-6918},
  url = {http://dx.doi.org/10.1142/S0219749908004195},
  DOI = {10.1142/s0219749908004195},
  number = {04},
  journal = {International Journal of Quantum Information},
  publisher = {World Scientific Pub Co Pte Ltd},
  author = {Chen,  XIU-BO and Wen,  QIAO-YAN and Guo,  FEN-ZHUO and Sun,  YING and Xu,  Gang and Zhu,  FU-CHEN},
  year = {2008},
  month = aug,
  pages = {899–906}
}

@article{Biham2002,
  title = {Grover's quantum search algorithm for an arbitrary initial mixed state},
  author = {Biham, Eli and Kenigsberg, Dan},
  journal = {Phys. Rev. A},
  volume = {66},
  issue = {6},
  pages = {062301},
  numpages = {4},
  year = {2002},
  month = {Dec},
  publisher = {American Physical Society},
  doi = {10.1103/PhysRevA.66.062301},
  url = {https://link.aps.org/doi/10.1103/PhysRevA.66.062301}
}

@article{Shapira2005,
  title = {Algebraic analysis of quantum search with pure and mixed states},
  author = {Shapira, Daniel and Shimoni, Yishai and Biham, Ofer},
  journal = {Phys. Rev. A},
  volume = {71},
  issue = {4},
  pages = {042320},
  numpages = {8},
  year = {2005},
  month = {Apr},
  publisher = {American Physical Society},
  doi = {10.1103/PhysRevA.71.042320},
  url = {https://link.aps.org/doi/10.1103/PhysRevA.71.042320}
}

@Inbook{Greenberger1989,
author="Greenberger, Daniel M.
and Horne, Michael A.
and Zeilinger, Anton",
editor="Kafatos, Menas",
title="Going Beyond Bell's Theorem",
bookTitle="Bell's Theorem, Quantum Theory and Conceptions of the Universe",
year="1989",
publisher="Springer Netherlands",
address="Dordrecht",
pages="69--72",
isbn="978-94-017-0849-4",
doi="10.1007/978-94-017-0849-4_10",
url="https://doi.org/10.1007/978-94-017-0849-4_10"
}

@article{Saffman2010,
  title = {Quantum information with Rydberg atoms},
  author = {Saffman, M. and Walker, T. G. and M\o{}lmer, K.},
  journal = {Rev. Mod. Phys.},
  volume = {82},
  issue = {3},
  pages = {2313--2363},
  numpages = {0},
  year = {2010},
  month = {Aug},
  publisher = {American Physical Society},
  doi = {10.1103/RevModPhys.82.2313},
  url = {https://link.aps.org/doi/10.1103/RevModPhys.82.2313}
}

@article{Doria2011,
  title = {Optimal Control Technique for Many-Body Quantum Dynamics},
  author = {Doria, Patrick and Calarco, Tommaso and Montangero, Simone},
  journal = {Phys. Rev. Lett.},
  volume = {106},
  issue = {19},
  pages = {190501},
  numpages = {4},
  year = {2011},
  month = {May},
  publisher = {American Physical Society},
  doi = {10.1103/PhysRevLett.106.190501},
  url = {https://link.aps.org/doi/10.1103/PhysRevLett.106.190501}
}

@article{Lloyd2014,
  title = {Information Theoretical Analysis of Quantum Optimal Control},
  author = {Lloyd, S. and Montangero, S.},
  journal = {Phys. Rev. Lett.},
  volume = {113},
  issue = {1},
  pages = {010502},
  numpages = {5},
  year = {2014},
  month = {Jul},
  publisher = {American Physical Society},
  doi = {10.1103/PhysRevLett.113.010502},
  url = {https://link.aps.org/doi/10.1103/PhysRevLett.113.010502}
}

@article{Gaetan2009,
    author = {Gaëtan, Alpha
and Miroshnychenko, Yevhen and Wilk, Tatjana and Chotia, Amodsen
and Viteau, Matthieu and Comparat, Daniel and Pillet, Pierre
and Browaeys, Antoine and Grangier, Philippe},
    title = {Observation of collective excitation of two individual atoms in the Rydberg blockade regime},
    journal = {Nature Physics},
    year = {2009},
    volume={5},
    pages={115},
    url={https://www.nature.com/articles/nphys1183},
    doi={10.1038/nphys1183}
}

@article{Campostrini2015b,
  title = {Quantum transitions driven by one-bond defects in quantum Ising rings},
  author = {Campostrini, Massimo and Pelissetto, Andrea and Vicari, Ettore},
  journal = {Phys. Rev. E},
  volume = {91},
  issue = {4},
  pages = {042123},
  numpages = {5},
  year = {2015},
  month = {Apr},
  publisher = {American Physical Society},
  doi = {10.1103/PhysRevE.91.042123},
  url = {https://link.aps.org/doi/10.1103/PhysRevE.91.042123}
}

@article{Odavic2023,
  title = {Complexity of frustration: A new source of non-local non-stabilizerness},
  volume = {15},
  ISSN = {2542-4653},
  url = {http://dx.doi.org/10.21468/SciPostPhys.15.4.131},
  DOI = {10.21468/scipostphys.15.4.131},
  number = {4},
  journal = {SciPost Physics},
  publisher = {Stichting SciPost},
  author = {Odavić,  Jovan and Haug,  Tobias and Torre,  Gianpaolo and Hamma,  Alioscia and Franchini,  Fabio and Giampaolo,  Salvatore Marco},
  year = {2023},
  month = oct 
}

@article{CIESLINSKI20241,
title = {Analysing quantum systems with randomised measurements},
journal = {Physics Reports},
volume = {1095},
pages = {1-48},
year = {2024},
note = {Analysing quantum systems with randomised measurements},
issn = {0370-1573},
doi = {https://doi.org/10.1016/j.physrep.2024.09.009},
url = {https://www.sciencedirect.com/science/article/pii/S0370157324003326},
author = {Paweł Cieśliński and Satoya Imai and Jan Dziewior and Otfried Gühne and Lukas Knips and Wiesław Laskowski and Jasmin Meinecke and Tomasz Paterek and Tamás Vértesi}
}

@Article{Elben2023,
author={Elben, Andreas
and Flammia, Steven T.
and Huang, Hsin-Yuan
and Kueng, Richard
and Preskill, John
and Vermersch, Beno{\^i}t
and Zoller, Peter},
title={The randomized measurement toolbox},
journal={Nature Reviews Physics},
year={2023},
month={Jan},
day={01},
volume={5},
number={1},
pages={9-24},
issn={2522-5820},
doi={10.1038/s42254-022-00535-2},
url={https://doi.org/10.1038/s42254-022-00535-2}
}

@Article{Huang2020,
author={Huang, Hsin-Yuan
and Kueng, Richard
and Preskill, John},
title={Predicting many properties of a quantum system from very few measurements},
journal={Nature Physics},
year={2020},
month={Oct},
day={01},
volume={16},
number={10},
pages={1050-1057},
issn={1745-2481},
doi={10.1038/s41567-020-0932-7},
url={https://doi.org/10.1038/s41567-020-0932-7}
}

@BOOK{MacKay2003,
  title     = {Information theory, inference and learning algorithms},
  author    = {MacKay, David J C},
  publisher = {Cambridge University Press},
  year      = {2003},
  address   = {Cambridge, England},
}

@inproceedings{Haah2016,
author = {Haah, Jeongwan and Harrow, Aram W. and Ji, Zhengfeng and Wu, Xiaodi and Yu, Nengkun},
title = {Sample-optimal tomography of quantum states},
year = {2016},
isbn = {9781450341325},
publisher = {Association for Computing Machinery},
address = {New York, NY, USA},
url = {https://doi.org/10.1145/2897518.2897585},
doi = {10.1145/2897518.2897585},
booktitle = {Proceedings of the Forty-Eighth Annual ACM Symposium on Theory of Computing},
pages = {913–925},
numpages = {13},
location = {Cambridge, MA, USA},
series = {STOC '16}
}

@misc{kozlej2025adiabaticstatepreparationthermalization,
      title={Adiabatic state preparation and thermalization of simulated phase noise in a Rydberg spin Hamiltonian}, 
      author={Tomas Kozlej and Gerard Pelegri and Jonathan D. Pritchard and Andrew J. Daley},
      year={2025},
      eprint={2505.04595},
      archivePrefix={arXiv},
      primaryClass={quant-ph},
      url={https://arxiv.org/abs/2505.04595}, 
}

@article{PhysRevA.97.053803,
  title = {Analysis of imperfections in the coherent optical excitation of single atoms to Rydberg states},
  author = {de L\'es\'eleuc, Sylvain and Barredo, Daniel and Lienhard, Vincent and Browaeys, Antoine and Lahaye, Thierry},
  journal = {Phys. Rev. A},
  volume = {97},
  issue = {5},
  pages = {053803},
  numpages = {9},
  year = {2018},
  month = {May},
  publisher = {American Physical Society},
  doi = {10.1103/PhysRevA.97.053803},
  url = {https://link.aps.org/doi/10.1103/PhysRevA.97.053803}
}

@article{fang2024probingcriticalphenomenaopen,
author = {Fang Fang  and Kenneth Wang  and Vincent S. Liu  and Yu Wang  and Ryan Cimmino  and Julia Wei  and Marcus Bintz  and Avery Parr  and Jack Kemp  and Kang-Kuen Ni  and Norman Y. Yao },
title = {Probing critical phenomena in open quantum systems using atom arrays},
journal = {Science},
volume = {390},
number = {6773},
pages = {601-605},
year = {2025},
doi = {10.1126/science.adq0278},
URL = {https://www.science.org/doi/abs/10.1126/science.adq0278},
eprint = {https://www.science.org/doi/pdf/10.1126/science.adq0278},}

@article{dag2024emergentdisordersubballisticdynamics,
  title = {Emergent Disorder and Sub-ballistic Dynamics in Quantum Simulations of the Ising Model Using Rydberg Atom Arrays},
  author = {Da\ifmmode \breve{g}\else \u{g}\fi{}, Ceren B. and Ma, Hanzhen and Eugenio, P. Myles and Fang, Fang and Yelin, Susanne F.},
  journal = {Phys. Rev. Lett.},
  volume = {135},
  issue = {25},
  pages = {250403},
  numpages = {7},
  year = {2025},
  month = {Dec},
  publisher = {American Physical Society},
  doi = {10.1103/jr7l-2cfb},
  url = {https://link.aps.org/doi/10.1103/jr7l-2cfb}
}

@article{Marcuzzi2017,
  title = {Facilitation Dynamics and Localization Phenomena in Rydberg Lattice Gases with Position Disorder},
  author = {Marcuzzi, Matteo and Min\'a\ifmmode \check{r}\else \v{r}\fi{}, Ji\ifmmode \check{r}\else \v{r}\fi{}\'{\i} and Barredo, Daniel and de L\'es\'eleuc, Sylvain and Labuhn, Henning and Lahaye, Thierry and Browaeys, Antoine and Levi, Emanuele and Lesanovsky, Igor},
  journal = {Phys. Rev. Lett.},
  volume = {118},
  issue = {6},
  pages = {063606},
  numpages = {6},
  year = {2017},
  month = {Feb},
  publisher = {American Physical Society},
  doi = {10.1103/PhysRevLett.118.063606},
  url = {https://link.aps.org/doi/10.1103/PhysRevLett.118.063606}
}

@article{bernien2017,
  title = {Probing many-body dynamics on a 51-atom quantum simulator},
  volume = {551},
  ISSN = {1476-4687},
  url = {http://dx.doi.org/10.1038/nature24622},
  DOI = {10.1038/nature24622},
  number = {7682},
  journal = {Nature},
  publisher = {Springer Science and Business Media LLC},
  author = {Bernien,  Hannes and Schwartz,  Sylvain and Keesling,  Alexander and Levine,  Harry and Omran,  Ahmed and Pichler,  Hannes and Choi,  Soonwon and Zibrov,  Alexander S. and Endres,  Manuel and Greiner,  Markus and Vuletić,  Vladan and Lukin,  Mikhail D.},
  year = {2017},
  month = nov,
  pages = {579–584}
}

@article{labuhn2016,
  title = {Tunable two-dimensional arrays of single Rydberg atoms for realizing quantum Ising models},
  volume = {534},
  ISSN = {1476-4687},
  url = {http://dx.doi.org/10.1038/nature18274},
  DOI = {10.1038/nature18274},
  number = {7609},
  journal = {Nature},
  publisher = {Springer Science and Business Media LLC},
  author = {Labuhn,  Henning and Barredo,  Daniel and Ravets,  Sylvain and de Léséleuc,  Sylvain and Macrì,  Tommaso and Lahaye,  Thierry and Browaeys,  Antoine},
  year = {2016},
  month = jun,
  pages = {667–670}
}

@article{bluvstein2022,
  title = {A quantum processor based on coherent transport of entangled atom arrays},
  volume = {604},
  ISSN = {1476-4687},
  url = {http://dx.doi.org/10.1038/s41586-022-04592-6},
  DOI = {10.1038/s41586-022-04592-6},
  number = {7906},
  journal = {Nature},
  publisher = {Springer Science and Business Media LLC},
  author = {Bluvstein,  Dolev and Levine,  Harry and Semeghini,  Giulia and Wang,  Tout T. and Ebadi,  Sepehr and Kalinowski,  Marcin and Keesling,  Alexander and Maskara,  Nishad and Pichler,  Hannes and Greiner,  Markus and Vuletić,  Vladan and Lukin,  Mikhail D.},
  year = {2022},
  month = apr,
  pages = {451–456}
}

@article{daug2016,
  title = {Multiatom Quantum Coherences in Micromasers as Fuel for Thermal and Nonthermal Machines},
  volume = {18},
  ISSN = {1099-4300},
  url = {http://dx.doi.org/10.3390/e18070244},
  DOI = {10.3390/e18070244},
  number = {7},
  journal = {Entropy},
  publisher = {MDPI AG},
  author = {Dağ,  Ceren and Niedenzu,  Wolfgang and M\"{u}stecaplıoğlu,  \"{O}zg\"{u}r and Kurizki,  Gershon},
  year = {2016},
  month = jun,
  pages = {244}
}

@article{Omran_2019,
   title={Generation and manipulation of Schrödinger cat states in Rydberg atom arrays},
   volume={365},
   ISSN={1095-9203},
   url={http://dx.doi.org/10.1126/science.aax9743},
   DOI={10.1126/science.aax9743},
   number={6453},
   journal={Science},
   publisher={American Association for the Advancement of Science (AAAS)},
   author={Omran, A. and Levine, H. and Keesling, A. and Semeghini, G. and Wang, T. T. and Ebadi, S. and Bernien, H. and Zibrov, A. S. and Pichler, H. and Choi, S. and Cui, J. and Rossignolo, M. and Rembold, P. and Montangero, S. and Calarco, T. and Endres, M. and Greiner, M. and Vuletić, V. and Lukin, M. D.},
   year={2019},
   month=aug, pages={570–574} }

@article{grafe2014,
  title = {On-chip generation of high-order single-photon W-states},
  volume = {8},
  ISSN = {1749-4893},
  url = {http://dx.doi.org/10.1038/nphoton.2014.204},
  DOI = {10.1038/nphoton.2014.204},
  number = {10},
  journal = {Nature Photonics},
  publisher = {Springer Science and Business Media LLC},
  author = {Gr\"{a}fe,  Markus and Heilmann,  René and Perez-Leija,  Armando and Keil,  Robert and Dreisow,  Felix and Heinrich,  Matthias and Moya-Cessa,  Hector and Nolte,  Stefan and Christodoulides,  Demetrios N. and Szameit,  Alexander},
  year = {2014},
  month = aug,
  pages = {791–795}
}

@article{Eibl2004,
  title = {Experimental Realization of a Three-Qubit Entangled $W$ State},
  author = {Eibl, Manfred and Kiesel, Nikolai and Bourennane, Mohamed and Kurtsiefer, Christian and Weinfurter, Harald},
  journal = {Phys. Rev. Lett.},
  volume = {92},
  issue = {7},
  pages = {077901},
  numpages = {4},
  year = {2004},
  month = {Feb},
  publisher = {American Physical Society},
  doi = {10.1103/PhysRevLett.92.077901},
  url = {https://link.aps.org/doi/10.1103/PhysRevLett.92.077901}
}

@article{roos2004,
  title = {Control and Measurement of Three-Qubit Entangled States},
  volume = {304},
  ISSN = {1095-9203},
  url = {http://dx.doi.org/10.1126/science.1097522},
  DOI = {10.1126/science.1097522},
  number = {5676},
  journal = {Science},
  publisher = {American Association for the Advancement of Science (AAAS)},
  author = {Roos,  Christian F. and Riebe,  Mark and Häffner,  Hartmut and Hänsel,  Wolfgang and Benhelm,  Jan and Lancaster,  Gavin P. T. and Becher,  Christoph and Schmidt-Kaler,  Ferdinand and Blatt,  Rainer},
  year = {2004},
  month = jun,
  pages = {1478–1480}
}

@article{kim2010,
  title = {Quantum simulation of frustrated Ising spins with trapped ions},
  volume = {465},
  ISSN = {1476-4687},
  url = {http://dx.doi.org/10.1038/nature09071},
  DOI = {10.1038/nature09071},
  number = {7298},
  journal = {Nature},
  publisher = {Springer Science and Business Media LLC},
  author = {Kim,  K. and Chang,  M.-S. and Korenblit,  S. and Islam,  R. and Edwards,  E. E. and Freericks,  J. K. and Lin,  G.-D. and Duan,  L.-M. and Monroe,  C.},
  year = {2010},
  month = jun,
  pages = {590–593}
}

@article{haffner2005,
  title = {Scalable multiparticle entanglement of trapped ions},
  volume = {438},
  ISSN = {1476-4687},
  url = {http://dx.doi.org/10.1038/nature04279},
  DOI = {10.1038/nature04279},
  number = {7068},
  journal = {Nature},
  publisher = {Springer Science and Business Media LLC},
  author = {H\"{a}ffner,  H. and H\"{a}nsel,  W. and Roos,  C. F. and Benhelm,  J. and Chek-al-kar,  D. and Chwalla,  M. and K\"{o}rber,  T. and Rapol,  U. D. and Riebe,  M. and Schmidt,  P. O. and Becher,  C. and G\"{u}hne,  O. and D\"{u}r,  W. and Blatt,  R.},
  year = {2005},
  month = dec,
  pages = {643–646}
}

@misc{Catalano2024_magic,
      title={Magic phase transition and non-local complexity in generalized $W$ State}, 
      author={A. G. Catalano and J. Odavić and G. Torre and A. Hamma and F. Franchini and S. M. Giampaolo},
      year={2024},
      eprint={2406.19457},
      archivePrefix={arXiv},
      primaryClass={quant-ph},
      url={https://arxiv.org/abs/2406.19457}, 
}

@article{Maric2020_induced,
  title = {Quantum phase transition induced by topological frustration},
  volume = {3},
  ISSN = {2399-3650},
  url = {http://dx.doi.org/10.1038/s42005-020-00486-z},
  DOI = {10.1038/s42005-020-00486-z},
  number = {1},
  journal = {Communications Physics},
  publisher = {Springer Science and Business Media LLC},
  author = {Marić,  Vanja and Giampaolo,  Salvatore Marco and Franchini,  Fabio},
  year = {2020},
  month = dec 
}

@article{Maric2020_destroy,
  title = {The frustration of being odd: how boundary conditions can destroy local order},
  volume = {22},
  ISSN = {1367-2630},
  url = {http://dx.doi.org/10.1088/1367-2630/aba064},
  DOI = {10.1088/1367-2630/aba064},
  number = {8},
  journal = {New Journal of Physics},
  publisher = {IOP Publishing},
  author = {Marić,  Vanja and Giampaolo,  Salvatore Marco and Kuić,  Domagoj and Franchini,  Fabio},
  year = {2020},
  month = aug,
  pages = {083024}
}

@misc{GeimArxiv2026,
      title={Engineering quantum criticality and dynamics on an analog-digital simulator}, 
      author={Alexandra A. Geim and Nazli Ugur Koyluoglu and Simon J. Evered and Rahul Sahay and Sophie H. Li and Muqing Xu and Dolev Bluvstein and Nik O. Gjonbalaj and Nishad Maskara and Marcin Kalinowski and Tom Manovitz and Ruben Verresen and Susanne F. Yelin and Johannes Feldmeier and Markus Greiner and Vladan Vuletic and Mikhail D. Lukin},
      year={2026},
      eprint={2602.18555},
      archivePrefix={arXiv},
      primaryClass={quant-ph},
      url={https://arxiv.org/abs/2602.18555}, 
}

@article{Buhrman2024,
  doi = {10.22331/q-2024-12-09-1552},
  url = {https://doi.org/10.22331/q-2024-12-09-1552},
  title = {State preparation by shallow circuits using feed forward},
  author = {Buhrman, Harry and Folkertsma, Marten and Loff, Bruno and Neumann, Niels M. P.},
  journal = {{Quantum}},
  issn = {2521-327X},
  publisher = {{Verein zur F{\"{o}}rderung des Open Access Publizierens in den Quantenwissenschaften}},
  volume = {8},
  pages = {1552},
  month = dec,
  year = {2024}
}

@article{Piroli2024,
  title = {Approximating Many-Body Quantum States with Quantum Circuits and Measurements},
  author = {Piroli, Lorenzo and Styliaris, Georgios and Cirac, J. Ignacio},
  journal = {Phys. Rev. Lett.},
  volume = {133},
  issue = {23},
  pages = {230401},
  numpages = {7},
  year = {2024},
  month = {Dec},
  publisher = {American Physical Society},
  doi = {10.1103/PhysRevLett.133.230401},
  url = {https://link.aps.org/doi/10.1103/PhysRevLett.133.230401}
}

@misc{farrell2025,
      title={Digital quantum simulations of scattering in quantum field theories using W states}, 
      author={Roland C. Farrell and Nikita A. Zemlevskiy and Marc Illa and John Preskill},
      year={2025},
      eprint={2505.03111},
      archivePrefix={arXiv},
      primaryClass={quant-ph},
      url={https://arxiv.org/abs/2505.03111}, 
}

@article{Catalano2024_battery,
  title = {Frustrating Quantum Batteries},
  volume = {5},
  ISSN = {2691-3399},
  url = {http://dx.doi.org/10.1103/PRXQuantum.5.030319},
  DOI = {10.1103/prxquantum.5.030319},
  number = {3},
  journal = {PRX Quantum},
  publisher = {American Physical Society (APS)},
  author = {Catalano,  A. G. and Giampaolo,  S. M. and Morsch,  O. and Giovannetti,  V. and Franchini,  F.},
  year = {2024},
  month = jul 
}

@misc{Catalano2024_TEBD,
  doi = {10.48550/ARXIV.2402.05198},
  url = {https://arxiv.org/abs/2402.05198},
  author = {Catalano,  Alberto Giuseppe},
  title = {Numerically efficient unitary evolution for Hamiltonians beyond nearest-neighbors},
  publisher = {arXiv},
  year = {2024},
  copyright = {arXiv.org perpetual,  non-exclusive license}
}

@article{Dong2016,
  title = {The a-cycle problem for transverse Ising ring},
  volume = {2016},
  ISSN = {1742-5468},
  url = {http://dx.doi.org/10.1088/1742-5468/2016/11/113102},
  DOI = {10.1088/1742-5468/2016/11/113102},
  number = {11},
  journal = {Journal of Statistical Mechanics: Theory and Experiment},
  publisher = {IOP Publishing},
  author = {Dong,  Jian-Jun and Li,  Peng and Chen,  Qi-Hui},
  year = {2016},
  month = nov,
  pages = {113102}
}

@article{Bravy2021,
  title = {Mitigating measurement errors in multiqubit experiments},
  author = {Bravyi, Sergey and Sheldon, Sarah and Kandala, Abhinav and Mckay, David C. and Gambetta, Jay M.},
  journal = {Phys. Rev. A},
  volume = {103},
  issue = {4},
  pages = {042605},
  numpages = {12},
  year = {2021},
  month = {Apr},
  publisher = {American Physical Society},
  doi = {10.1103/PhysRevA.103.042605},
  url = {https://link.aps.org/doi/10.1103/PhysRevA.103.042605}
}

@misc{Aquila,
  doi = {10.48550/ARXIV.2306.11727},
  url = {https://arxiv.org/abs/2306.11727},
  author = {Wurtz,  Jonathan and Bylinskii,  Alexei and Braverman,  Boris and Amato-Grill,  Jesse and Cantu,  Sergio H. and Huber,  Florian and Lukin,  Alexander and Liu,  Fangli and Weinberg,  Phillip and Long,  John and Wang,  Sheng-Tao and Gemelke,  Nathan and Keesling,  Alexander},
  title = {Aquila: QuEra's 256-qubit neutral-atom quantum computer},
  publisher = {arXiv},
  year = {2023},
  copyright = {arXiv.org perpetual,  non-exclusive license}
}

@misc{Gioia2023,
  doi = {10.48550/ARXIV.2310.10716},
  url = {https://arxiv.org/abs/2310.10716},
  author = {Gioia,  Lei and Thorngren,  Ryan},
  title = {$W$ state is not the unique ground state of any local Hamiltonian},
  publisher = {arXiv},
  year = {2023},
  copyright = {Creative Commons Attribution 4.0 International}
}

@misc{Lukin2024,
  doi = {10.48550/ARXIV.2405.21019},
  url = {https://arxiv.org/abs/2405.21019},
  author = {Lukin,  Alexander and Schiffer,  Benjamin F. and Braverman,  Boris and Cantu,  Sergio H. and Huber,  Florian and Bylinskii,  Alexei and Amato-Grill,  Jesse and Maskara,  Nishad and Cain,  Madelyn and Wild,  Dominik S. and Samajdar,  Rhine and Lukin,  Mikhail D.},
  title = {Quantum quench dynamics as a shortcut to adiabaticity},
  publisher = {arXiv},
  year = {2024},
  copyright = {arXiv.org perpetual,  non-exclusive license}
}

@article{Cruz2019,
  title = {Efficient Quantum Algorithms for GHZ and W States,  and Implementation on the IBM Quantum Computer},
  volume = {2},
  ISSN = {2511-9044},
  url = {http://dx.doi.org/10.1002/qute.201900015},
  DOI = {10.1002/qute.201900015},
  number = {5–6},
  journal = {Advanced Quantum Technologies},
  publisher = {Wiley},
  author = {Cruz,  Diogo and Fournier,  Romain and Gremion,  Fabien and Jeannerot,  Alix and Komagata,  Kenichi and Tosic,  Tara and Thiesbrummel,  Jarla and Chan,  Chun Lam and Macris,  Nicolas and Dupertuis,  Marc‐André and Javerzac‐Galy,  Clément},
  year = {2019},
  month = apr 
}

@book{Meystre2007,
    author = { Meystre, Pierre and Sargent Murray } ,
    title = {Elements of Quantum Optics},
    publisher ={Springer Berlin, Heidelberg} ,
    year = {2007},
    doi = {10.1007/978-3-540-74211-1}
}

@article{efron1979,
  title     = {Bootstrap methods: Another look at the jackknife},
  author    = {Efron, Bradley},
  journal   = {The Annals of Statistics},
  volume    = {7},
  number    = {1},
  pages     = {1--26},
  year      = {1979},
  publisher = {Institute of Mathematical Statistics},
  doi       = {10.1214/aos/1176344552}
}

@book{efron1994,
  title     = {An Introduction to the Bootstrap},
  author    = {Efron, Bradley and Tibshirani, Robert J.},
  publisher = {Chapman and Hall/CRC},
  year      = {1994},
  address   = {New York},
  isbn      = {978-0412042317},
  doi = {10.1201/9780429246593}
}

@inbook{Toulouse1986,
  title = {Theory of the frustration effect in spin glasses: I},
  ISSN = {1793-1436},
  url = {http://dx.doi.org/10.1142/9789812799371_0009},
  DOI = {10.1142/9789812799371_0009},
  booktitle = {Spin Glass Theory and Beyond},
  publisher = {WORLD SCIENTIFIC},
  author = {Toulouse,  G.},
  year = {1986},
  month = nov,
  pages = {99–103}
}

@misc{Talath2025,
  doi = {10.48550/ARXIV.2507.02849},
  url = {https://arxiv.org/abs/2507.02849},
  author = {Talath,  H. and Govindaraja,  B. P. and Divyamani,  B. G. and H.,  Akshata Shenoy and Devi,  A. R. Usha and {Sudha}},
  title = {Three-qubit W state tomography via full and marginal state reconstructions on ibm\_osaka},
  publisher = {arXiv},
  year = {2025},
  copyright = {Creative Commons Attribution 4.0 International}
}

@article{Granade2016,
  title = {Practical Bayesian tomography},
  volume = {18},
  ISSN = {1367-2630},
  url = {http://dx.doi.org/10.1088/1367-2630/18/3/033024},
  DOI = {10.1088/1367-2630/18/3/033024},
  number = {3},
  journal = {New Journal of Physics},
  publisher = {IOP Publishing},
  author = {Granade,  Christopher and Combes,  Joshua and Cory,  D G},
  year = {2016},
  month = mar,
  pages = {033024}
}

@article{Blume2010,
  title = {Optimal,  reliable estimation of quantum states},
  volume = {12},
  ISSN = {1367-2630},
  url = {http://dx.doi.org/10.1088/1367-2630/12/4/043034},
  DOI = {10.1088/1367-2630/12/4/043034},
  number = {4},
  journal = {New Journal of Physics},
  publisher = {IOP Publishing},
  author = {Blume-Kohout,  Robin},
  year = {2010},
  month = apr,
  pages = {043034}
}

@article{Schack2001,
  title = {Quantum Bayes rule},
  volume = {64},
  ISSN = {1094-1622},
  url = {http://dx.doi.org/10.1103/PhysRevA.64.014305},
  DOI = {10.1103/physreva.64.014305},
  number = {1},
  journal = {Physical Review A},
  publisher = {American Physical Society (APS)},
  author = {Schack,  R\"{u}diger and Brun,  Todd A. and Caves,  Carlton M.},
  year = {2001},
  month = jun 
}

@article{Lukens2020,
  title = {A practical and efficient approach for Bayesian quantum state estimation},
  volume = {22},
  ISSN = {1367-2630},
  url = {http://dx.doi.org/10.1088/1367-2630/ab8efa},
  DOI = {10.1088/1367-2630/ab8efa},
  number = {6},
  journal = {New Journal of Physics},
  publisher = {IOP Publishing},
  author = {Lukens,  Joseph M and Law,  Kody J H and Jasra,  Ajay and Lougovski,  Pavel},
  year = {2020},
  month = jun,
  pages = {063038}
}

@book{Collura2024,
  title = {Tensor Network Techniques for Quantum Computation},
  ISBN = {9788898587049},
  url = {http://dx.doi.org/10.22323/9788898587049},
  DOI = {10.22323/9788898587049},
  publisher = {SISSA Medialab s.r.l.},
  author = {Collura,  Mario and Lami,  Guglielmo and Ranabhat,  Nishan and Santini,  Alessandro},
  year = {2024},
  month = dec 
}

@article{Paeckel2019,
  title = {Time-evolution methods for matrix-product states},
  volume = {411},
  ISSN = {0003-4916},
  url = {http://dx.doi.org/10.1016/j.aop.2019.167998},
  DOI = {10.1016/j.aop.2019.167998},
  journal = {Annals of Physics},
  publisher = {Elsevier BV},
  author = {Paeckel,  Sebastian and K\"{o}hler,  Thomas and Swoboda,  Andreas and Manmana,  Salvatore R. and Schollw\"{o}ck,  Ulrich and Hubig,  Claudius},
  year = {2019},
  month = dec,
  pages = {167998}
}

@book{Cover2006,
  title     = {Elements of Information Theory},
  author    = {Cover, Thomas M. and Thomas, Joy A.},
  year      = {2006},
  edition   = {2},
  publisher = {Wiley-Interscience},
  address   = {Hoboken, NJ, USA},
  isbn      = {978-0-471-24195-9}
}

@article{Ebert2015,
  title = {Coherence and Rydberg Blockade of Atomic Ensemble Qubits},
  volume = {115},
  ISSN = {1079-7114},
  url = {http://dx.doi.org/10.1103/PhysRevLett.115.093601},
  DOI = {10.1103/physrevlett.115.093601},
  number = {9},
  journal = {Physical Review Letters},
  publisher = {American Physical Society (APS)},
  author = {Ebert,  M. and Kwon,  M. and Walker,  T. G. and Saffman,  M.},
  year = {2015},
  month = aug 
}

@Misc{methods,
  note = {Materials and methods are available as supplementary material},
}

@article{Torre2025,
  title = {Interplay between local and non-local frustration in the 1D ANNNI chain: I. The even case},
  volume = {2025},
  ISSN = {1742-5468},
  url = {http://dx.doi.org/10.1088/1742-5468/adc897},
  DOI = {10.1088/1742-5468/adc897},
  number = {5},
  journal = {Journal of Statistical Mechanics: Theory and Experiment},
  publisher = {IOP Publishing},
  author = {Torre,  G and Catalano,  A G and Kožić,  S B and Franchini,  F and Giampaolo,  S M},
  year = {2025},
  month = may,
  pages = {053101}
}

@misc{Catalano2025_2D,
  doi = {10.48550/ARXIV.2509.03574},
  url = {https://arxiv.org/abs/2509.03574},
  author = {Catalano,  Alberto Giuseppe and Reinić,  Nora and Torre,  Gianpaolo and Kožić,  Sven Benjamin and Delić,  Karlo and Montangero,  Simone and Franchini,  Fabio and Giampaolo,  Salvatore Marco},
  title = {A new rung on the ladder: exploring topological frustration towards two dimensions},
  publisher = {arXiv},
  year = {2025},
  copyright = {arXiv.org perpetual,  non-exclusive license}
}
%TC:endignore

\end{document}